\documentstyle[12pt,epsf]{article}
\def\Vec#1{\mbox{\boldmath $ #1 $}}
\def\sVec#1{\mbox{\small\boldmath $ #1 $}}
\begin{document}
\begin{center}
\begin{Large}
Photonic band calculation in the form of $\Vec{k}(\omega)$ \\
including evanescent waves\\[5mm]
\end{Large}
K. Cho$^{(a)}$, J. Ushida$^{(b)}$, and M. Bamba$^{(a)}$ \\
a) Graduate School of Engineering Science, Osaka University, Toyonaka, 
560-8531 Japan\\
b) Fundamental Research Lab., NEC Corporation, 34 Miyukigaoka, Tsukuba,
305-8501 Japan\\[5mm]
\end{center}
Abstract

  We give a general method to calculate photonic band structure in the form 
of wave number $k$ as a function of frequency $\omega$, which is required 
whenever we want to calculate signal intensity related with photonic band 
structure.   This method is based on the fact that the elements of the 
coefficient matrix for the plane wave expansion of the Maxwell equations 
contain wave number up to the second order, which allows us to rewrite 
the original eigenvalue equation for $\omega^2$ into that for wave number. 
This method is much better, especially for complex wave numbers, than the 
transfer matrix method of Pendry, which gives the eigenvalues in the form 
of exp$[ikd]$ .   As a simplest example, we show a comparison of $\omega(k)$ 
and $k(\omega)$ for a model of intersecting square rods. \\[5mm]

\section{Introduction}

Usual band structure is calculated as eigen frequency for a given wave vector 
of a photonic crystal.   This is common to the electronic band structure 
of crystals, too.   The information about \{$\hbar\omega_{j}(\Vec{k})$\} 
serves as a fundamental quantity to discuss the behavior of the crystal. 
However, when we want to calculate any signals from this crystal, we have to 
work on a sample with boundary, and consider the relationship between the 
solutions in- and outside the sample.  For both photonic and electronic bands, 
we need to connect the solutions across the boundary, and in doing this 
we have to consider both propagating and evanescent wave components.  
For this purpose, we have to find out a method to solve the eigenvalue 
equation, not in the form of $\omega_{j}(k)$, but in the form of $k_{j}
(\omega)$ for real $\omega$. 

The eigenvalue equation can be expressed in general as a linear equations 
for the Fourier components $\{\Vec{X}\}$ of wave functions or EM field 
\begin{equation}
\label{eqn:a1}
  {\bf S} \Vec{X} = 0
\end{equation}
for both electronic and photonic band problems.   This is the Schr\"odinger 
equation for electronic problems, and Maxwell equations for photonic 
crystals.   The coefficient matrix is a function of ($\omega, k$) as well as 
Fourier components of ''potential'' term, which depend on the reciprocal lattice 
vectors.   This matrix has the form 
\begin{equation}
\label{eqn:a2}
  {\bf S} = {\bf S}' - \lambda {\bf 1}
\end{equation}
where $\lambda = \hbar\omega$ (energy) for Schr\"odinger equation, 
and $\lambda = \omega^2$ 
for Maxwell equations.   Thus the eigenvalues of the matrix ${\bf S}'$ 
give the band structure $\{\omega_{j}(\Vec{k})\}$ in a straightforward manner. 

On the other hand, it is not so clear to find a way to get $\Vec{k}_{j}
(\omega)$ from the same equation.   However, if we note the following 
simple argument, which is known for a numerical solution of eigenvalue 
equation, we are lead to an appropriate way to get $\Vec{k}_{j}(\omega)$. 

Let us consider a set of equations 
\begin{equation}
\label{eqn:a3}
  {\bf A} \ \Vec{X} = 0 \ .
\end{equation}
where the matrix {\bf A} is assumed to depend on a parameter $\lambda$.  
We now ask the condition for $\lambda$ to obtain a nontrivial 
solution $\Vec{X}$.  
If the $\lambda$-dependence of the matrix {\bf A} is 
\begin{equation}
\label{eqn:a4}
  {\bf A} = {\bf B} - \lambda\ {\bf C} \ , 
\end{equation} 
the condition for the nontrivial solution, i.e.,  
the eigenvalue equation for $\lambda$, is written as 
\begin{equation} 
\label{eqn:a5}
  {\rm det}|{\bf C}^{-1}\ {\bf B} - \lambda {\bf 1}| = 0 \ .
\end{equation}

In a similar way, for a matrix {\bf A} given in the following form, 
\begin{equation}
\label{eqn:a6}
  {\bf A} = {\bf B} + \lambda\ {\bf C} - \lambda^2\ {\bf D}\ , 
\end{equation} 
we rewrite eq.(\ref{eqn:a3}) by using 
new variables $\Vec{Y} = \lambda \Vec{X}$ as 
\begin{equation}
\label{eqn:a6x}
 {\bf D}^{-1}{\bf B} \Vec{X} +  {\bf D}^{-1}{\bf C} \Vec{Y} 
    - \lambda \Vec{Y} = 0
\end{equation}
Then, the condition to have nontrivial solution for $\{\Vec{X}, \Vec{Y}\}$, 
which works as an eigenvalue equation for $\lambda$, can be written 
as det$|{\bf U} - \lambda {\bf 1}|= 0$ , where 
\begin{equation}
\label{eqn:a7}
  {\bf U} =  
\left[ 
\begin{array}{rr}
 0\ ,   &   {\bf 1}  \\
{\bf D}^{-1}{\bf B}\ ,   &   {\bf D}^{-1}{\bf C} 
\end{array}
\right]
\end{equation}

This way of rewriting is possible for any power of $\lambda$, leading to  
the eigenvalue problem for $\lambda$.  Being well-known 
as a technique to get appropriate form of eigenvalue equation, it has been 
used for the calculation of multibranch polariton dispersion curves in the 
form of $k(\omega)$ \cite{polariton}.   

The cases of 
electronic and photonic band structure calculation in plane wave expansion 
have $\lambda \ ( = k )$ up to the second order.  This allows us to write down 
the general matrix equation for the arbitrary configuration as shown in detail 
below, where we will treat the case of photonic bands explicitly.  Its 
application to the case of electronic bands is straightforward, as mentioned 
above.  \\[5mm]

\section{Formulation} 

If we eliminate the electric field in the Maxwell equations for a photonic crystal ("H-method" \cite{H-method}), and make a Fourier series expansion with respect to the reciprocal lattice vectors $\Vec{G}$ of the crystal, we get a set of equations 
\begin{equation}
\label{eqn:a8}
  \sum_{\eta} \sum_{\sVec{G}'} S_{\sVec{G},\sVec{G}'}^{(\xi\eta)} \ 
                  H_{\sVec{k}, \sVec{G}'}^{(\eta)} = 0
\end{equation}
where ($\xi, \eta$) are the Cartesian coordinates, and 
\begin{equation}
\label{eqn:a9}
 {\bf S}_{\sVec{G},\sVec{G}'} = \omega^2 \mu_{\sVec{G}-\sVec{G}'} 
            + (1/\epsilon)_{\sVec{G}-\sVec{G}'}\ 
           (\Vec{k}+ \Vec{G}) \times (\Vec{k}+ \Vec{G}') \times    \ .
\end{equation}
In these expressions, $\epsilon$ and $\mu$ are the dielectric constant and 
magnetic permeability with periodic structure of the photonic crystal in 
consideration, and $\Vec{H}_{\sVec{k}, \sVec{G}'}$ the magnetic field of 
light in Fourier representation.   For a perfect crystal, $\Vec{k}$ is a good 
quantum number, and det$|{\bf S}| = 0$ gives the nontrivial solutions, 
i.e., the dispersion relation $\{\omega_{j}(\Vec{k})\}$ of the eigenmodes. 

It should be noted that the coefficient matrix ${\bf S}$ contains $\Vec{k}$ 
up to the second order.   Let us introduce a surface of this 
photonic crystal, which is always necessary to define an observation process. 
For simplicity we assume that the outside is vacuum. 
We take the $z$ axis to the surface normal direction (toward inside), and the incident light is within the $xz$ plane.   Then, the incident wave vector has components $(k_{\parallel}, 0, \kappa)$. If we denote the reciprocal lattice vectors of the surface periodic 2D lattice as $\{\Vec{g}\}$, there are reflected and transmitted light beams with the lateral wave vector components of 
 $\{\Vec{k}_{\parallel} + \Vec{g}\}$.  
The $z$-components of the wave vectors of reflected and transmitted lights 
should be determined so as they satisfy the dispersion relations in vacuum and 
bulk crystal, respectively.   
An important feature about the $z$-components is that 
they may well be complex numbers, namely, they can be evanescent waves.  Because of the presence of the surface, which breaks the translational symmetry, the 
evanescent solutions are allowed as long as their amplitudes decay as they go 
away from the surface. 

Noting the identity 
\begin{equation}
  (\Vec{k}+ \Vec{G}) \times (\Vec{k}+ \Vec{G}') \times \Vec{H}_{\sVec{k}, \sVec{G}'} = (\Vec{k}+ \Vec{G}')\ \{(\Vec{k}+ \Vec{G}) \cdot \Vec{H}_{\sVec{k}, \sVec{G}'}\} - \Vec{H}_{\sVec{k}, \sVec{G}'}\ \{(\Vec{k}+ \Vec{G})\cdot(\Vec{k}+ \Vec{G}')\}
\end{equation}
and 
\begin{equation}
 \Vec{k}+ \Vec{G} = (k_{\parallel}+G_{x})\hat{\rm e}_{x} + G_{y}\hat{\rm e}_{y} + (\kappa + G_{z})\hat{\rm e}_{z} \ ,
\end{equation}
we obtain the components of ${\bf S}$ for a given block 
of $(\Vec{G}, \Vec{G}')$ as 
\begin{equation} 
  {\bf S} = \omega^2 \mu_{\sVec{G}- \sVec{G}'} {\bf 1} 
     + \{- (\Vec{k}+\Vec{G})\cdot(\Vec{k}+\Vec{G}') {\bf 1} 
        + {\bf S}_{1} \} (1/\epsilon)_{\sVec{G}- \sVec{G}'}\ ,
\end{equation}
where 
\begin{eqnarray}
 {\bf S}_{1} = 
\left[
\begin{array}{rrr}
(k_{\parallel}+G_{x}')(k_{\parallel}+G_{x})\ ,&  (k_{\parallel}+G_{x}')G_{y}\ ,
         &   (k_{\parallel}+G_{x}')(\kappa + G_{z})\\
G_{y}'(k_{\parallel}+G_{x})\ , & G_{y}' G_{y}\ , & G_{y}'(\kappa + G_{z})\\
(\kappa + G_{z}')(k_{\parallel}+G_{x})\ , & (\kappa + G_{z}')G_{y}\ , & 
                             (\kappa + G_{z}') (\kappa + G_{z}) 
\end{array}
\right] \ .
\end{eqnarray}

Equation (\ref{eqn:a8}) should be supplemented by the condition  $\nabla\cdot
\Vec{H} = 0$, which requires the transversal nature of the magnetic field 
at any point.  In terms of the Fourier components, this condition leads, for 
all $\Vec{G}$'s, to 
\begin{equation}
 (k_{\parallel} + G_{x}') H_{\sVec{k}, \sVec{G}'}^{(x)} 
   + G_{y}' H_{\sVec{k}, \sVec{G}'}^{(y)}  + 
 (\kappa + G_{z}') H_{\sVec{k}, \sVec{G}'}^{(z)} = 0 . 
\end{equation}
These equations are used to eliminate $\{H_{\sVec{k}, \sVec{G}'}^{(x)}\}$ 
from eq.(\ref{eqn:a8}), which leads to the final form of the equations 
to be solved as 
\begin{equation}
 \sum_{\sVec{G}'}\ [\{ S^{(\xi y)} - \frac{G'_{y}}{k_{\parallel} + 
                  G'_{x}} S^{(\xi x)} \} H_{\sVec{k}, \sVec{G}'}^{(y)} 
               + \{ S^{(\xi z)} - \frac{\kappa + G'_{z}}{k_{\parallel} + 
                  G'_{x}} S^{(\xi x)} \} H_{\sVec{k}, \sVec{G}'}^{(z)}] = 0 \ .
\end{equation}
If we take N reciprocal lattice vectors, we need to determine 2N components 
$\{H_{\sVec{k}, \sVec{G}'}^{(y)},\ H_{\sVec{k}, \sVec{G}'}^{(z)}\}$. 
Choosing $y$ and $z$ for  $\xi$, we explicitly write   
this matrix equation in the form of eq.(\ref{eqn:a6}), which can be 
transformed into the eigenvalue problem for $\kappa$ as discussed above. 
The size of the matrix equation for $\kappa$ is $4N \times 4N$, which gives 
$2N$ solutions for each of the positive and negative directions of $\kappa$, 
including propagating and evanescent modes.  
The factor 2 corresponds to the polarization direction.  \\[5mm]

Fig.1 shows an example of comparison of the band calculations according 
to this method and H-method for a photonic bands of intersecting square rods 
for propagating modes.  The number of the reciprocal lattice vectors used 
for this calculation is $N = 9\times9\times9$.   As expected, the calculated 
bands as $\omega(\Vec{k})$  and $\Vec{k}(\omega)$ agrees very well.  Fig. 2 
shows the bands of complex wave vectors.   The branches with large imaginary 
part are obtained as smooth curves, indicating the stability of numerical 
calculation.   However, the fact that the Im$[\kappa]$ can exceed 
$2\pi \times 6$ for the present choice of $N$ indicates that in terms of 
the transfer matrix method we have to deal with the eigenvalues ranging 
from $10^{14}$ to $10^{-14}$. For higher frequency regions, we need shorter 
wavelength components, i.e., more reciprocal lattice vectors, which 
gives steeper evanescent wave components.  Thus the difficulty to solve 
eigenvalue problem of $\kappa$ increases.   In this way, the practical limit 
is reached rather quickly if we use the transfer matrix method.  The reason why we need more reciprocal lattice vectors for higher frequency is that the 
boundary conditions for a higher frequency light should be given at finer 
spatial mesh points, because it has shorter wavelength.

\section{Discussion} 

\subsection{Comparison with transfer matrix method} 

It is well known that the band calculation to get $\{\Vec{k}_{j}
(\omega)\}$ can be 
done via so-called transfer matrix method \cite{Pendry}.   In this method, 
one discretizes the Maxwell equations in an appropriate mesh, which provide 
the relationship among the EM field components ascribed to these mesh points. 
From this set of difference equations, one can derive 
a transfer matrix ${\bf T}$ which relates the EM fields at 
various positions on one lattice plane with those on the neighboring lattice 
plane.  Since the EM fields on the corresponding positions on the neighboring 
lattice planes (with a spacing $\Vec{d}$) are related according to 
Bloch theorem, the eigenvalues of ${\bf T}$ should be 
exp$[i\Vec{k}\cdot\Vec{d}]$, i.e., 
\begin{equation}
  {\rm det}|{\bf T} - \exp[i\Vec{k}\cdot\Vec{d}]| = 0 \ .
\end{equation}

This logic is certainly correct, but it is not quite convenient to carry it 
out as a numerical calculation.   Since we need all the possible solutions 
of the band states including evanescent waves, the eigenvalues 
$\exp[i\Vec{k}\cdot\Vec{d}]$ can be extremely large or small, which causes 
a serious trouble to the numerical calculation.   This means that the 
accuracy of calculating $\Vec{k}$ can become rather poor.   This has been 
encountered in our previous work to determine the branching ratio of an 
incident light into various waves in the bulk \cite{Minami}.  For lower 
frequency region, the result is not too bad.  However, as we examine higher 
frequency region, we start to see the frequencies where the reflectivity 
exceeds unity, i.e., a typical violation of physics principle.  
We believe that this is caused by the inaccuracy of calculating $\Vec{k}$'s 
for a given frequency.  

In contrast, the formulation in the previous section gives an eigenvalue 
equation for $\kappa(\omega)$, not for exp[i$\kappa d$].   Therefore, 
the range of eigenvalues is not tremendously large, as shown in Fig.2, 
and we can hope a good accuracy in their numerical values.

\subsection{Branching ratio in the reflected and transmitted light beams}

For a given incident light, there arise various beams of reflected and 
transmitted lights with propagating and evanescent character 
due to (i) the multibranch structure of photonic bands and 
(ii) the Bragg scattering from the surface periodicity.   In order to fix 
the amplitude ratio of these different beams, it is enough to consider the 
Maxwell boundary conditions (MBC's) at the surface.   
In spite of the multibranch structure of photonic bands, there is no need 
of ''additional boundary condition (ABC)'' \cite{ABC}, because we are dealing 
with macroscopic local response without any spatial dispersion effect. 

Since the surface is not uniform, we need to require MBC's at each point.  
Alternatively, we may require the MBC's for each Fourier components of 
electric and magnetic fields with respect to the surface reciprocal lattice 
vectors \{$\Vec{g}$\}.   The number of $\Vec{g}$'s to be used should be 
determined by the desired accuracy of numerical calculation.   For each 
$\Vec{g}$, we take the corresponding beams on both sides of the surface, 
and make them fulfil the MBC's.   In this way we can determine the amplitudes 
of all the beams uniquely.  

It should be noted that any beam in the photonic crystal has the form of 
Bloch wave, i.e.,  a product of $\Vec{k}$-dependent phase factor and lattice 
periodic part.   Therefore, its amplitude at the surface changes according 
to the position of a surface cross section for any given Miller indices.  This 
means that the beam amplitudes of transmitted and reflected lights depend on 
the surface cross section for any given Miller indices.  This was explicitly 
demonstrated by us for the [100] surface of the intersecting square rods, which shows remarkable difference among different cross sections of the surfaces \cite{Minami}.  This fact shows a possibility of designing the surface to control 
the flow of light energy in desired beams.  For this calculation also, the new 
method of calculating $\Vec{k}(\omega)$ should be quite useful, because the 
boundary conditions depend on the accuracy of evanescent waves. 

\subsection{Other applications}

In more realistic situations, we have to consider the finite size of 
photonic crystals, or even systems without periodicity.  In such cases, 
the method of FDTD (Finite Difference in Time Domain) method \cite{FDTD} 
is well known and used quite popularly.  The present method may be used 
in such cases, if we put the matter in a periodic system with 
super-cell construction.  In this way, we can hopefully extend the 
present scheme to a wide class of matter systems to calculate their 
optical responses in frequency domain in a straightforward way.  
This will provide us with a complimentary information from that of FDTD 
method. 

For the discussion of nonlinear processes in photonic crystals, 
we also need to know how much of an incident light is fed by a particular 
branch of photonic bands.   Thus, we need to prepare this input information 
by the argument of boundary conditions, and this is important both for 
fundamental and applicational purposes. \\[5mm]

\begin{flushleft}
\underline{Acknowledgments}\\
\end{flushleft}
This work was supported in part by the Grant-in-Aid for Scientific Research 
(No.15540311) of the Ministry of Education, Culture, Sports, Science and Technology of Japan.

\newpage
%\begin{flushleft}
%Figure captions \\
%\end{flushleft}
%\begin{enumerate}
%\item [Fig.1]: Band structure of an intersecting square-rod model 
%      calculated as        $\omega(\Vec{k})$ and $\Vec{k}(\omega)$ for real 
%      $\Vec{k} = (0.1, 0, \kappa) 2\pi/d$. The number of the reciprocal lattice 
%      points is $9\times9\times9$, and other parameters are the same as in 
%      ref.\cite{Minami}. \\
%\item [Fig.2]: Imaginary parts of the evanescent solutions of $\Vec{k}(\omega)$ 
%       for the same frequency range as in Fig.1.   The number of branches 
%       in this region is $2\times9\times9$, where the factor $2$ counts 
%       the polarization degree of freedom, including real (propagating) modes. 
%\end{enumerate}
%---------------------------------------------------------------------------
\begin{figure*}[tb]
 \begin{center}
  \epsfxsize=400pt \epsfbox{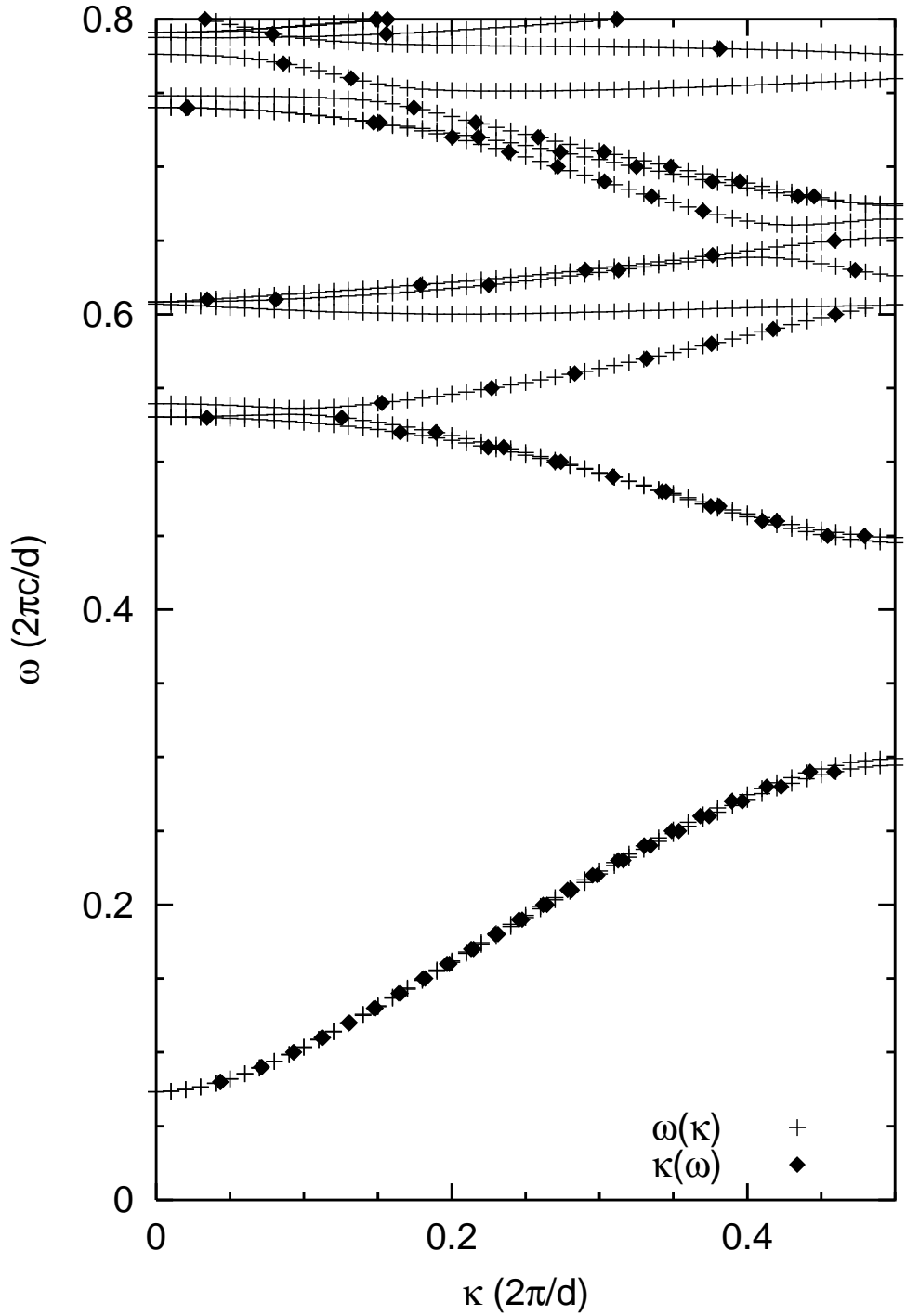}
  \end{center}
 \caption{
 Band structure of an intersecting square-rod model 
 calculated as          $\omega(\Vec{k})$ and $\Vec{k}(\omega)$ for real 
 $\Vec{k} = (0.1, 0, \kappa) 2\pi/d$. The number of the reciprocal lattice 
 points is $9\times9\times9$, and other parameters are the same as in 
 ref.\cite{Minami}. 
 }
\end{figure*}
%---------------------------------------------------------------------------
\begin{figure*}[tb]
 \begin{center}
  \epsfxsize=420pt  \epsfbox{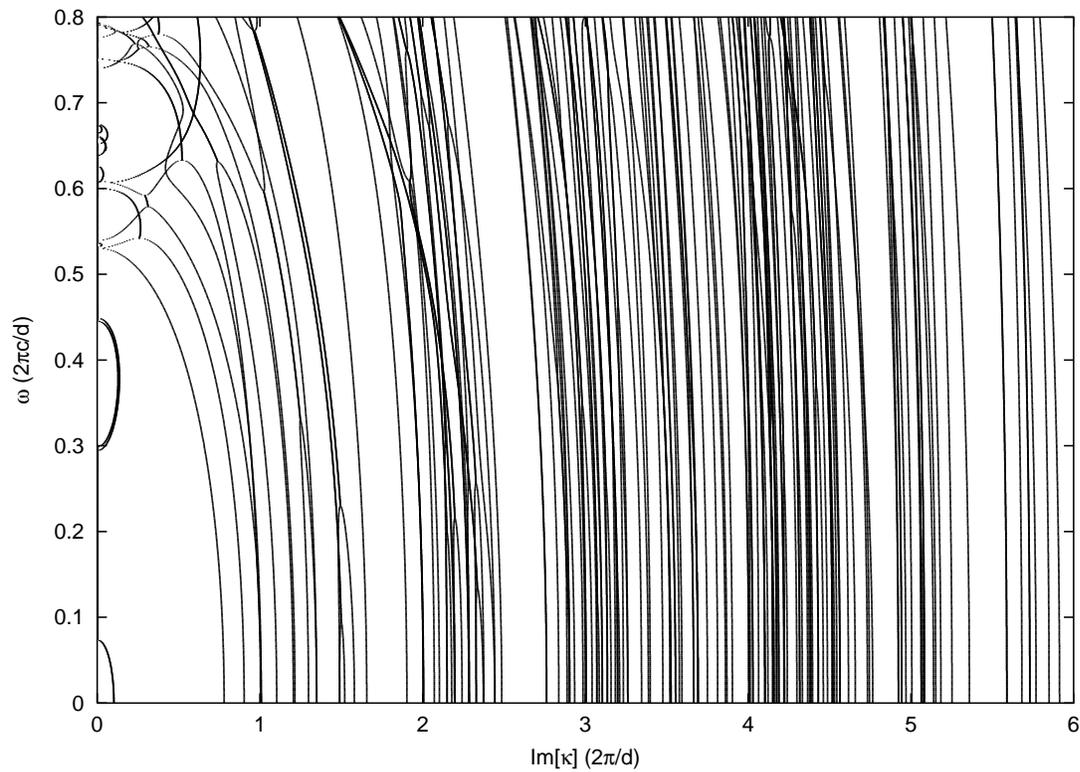}
 \end{center}
 \caption{
 Imaginary parts of the evanescent solutions of $\Vec{k}(\omega)$ 
 for the same frequency range as in Fig.1.   The number of branches 
 in this region is $2\times9\times9$, where the factor $2$ counts 
 the polarization degree of freedom, including real (propagating) modes. 
 }
\end{figure*}
%---------------------------------------------------------------------------
%
\end{document}